# Experimental evidence of large-gap two-dimensional topological insulator on the surface of ZrTe$_5$

R. Wu*,[1,2] J.-Z. Ma*,[1] L.-X. Zhao,[1] S.-M. Nie,[1] X. Huang,[1,2] J.-X. Yin,[1,2] B.-B. Fu,[1] P. Richard,[1,3] G.-F. Chen,[1,3] Z. Fang,[1,3] X. Dai,[1,3] H.-M. Weng,[1,3,†] T. Qian,[1,‡] H. Ding,[1,3] and S. H. Pan[1,2,3,§]

[1] *Beijing National Laboratory for Condensed Matter Physics and Institute of Physics, Chinese Academy of Sciences, Beijing 100190, China*

[2] *Department of Physics and Texas Center for Superconductivity, University of Houston, Houston, Texas 77204, USA*

[3] *Collaborative Innovation Center of Quantum Matter, Beijing, China*

* These authors contributed to this work equally.
Corresponding authors: hmweng@iphy.ac.cn, tqian@iphy.ac.cn, span@iphy.ac.cn

## Abstract

Two-dimensional (2D) topological insulators (TIs) with a large bulk band-gap are promising for experimental studies of the quantum spin Hall effect and for spintronic device applications. Despite considerable theoretical efforts in predicting large-gap 2D TI candidates, only few of them have been experimentally verified. Here, by combining scanning tunneling microscopy/spectroscopy and angle-resolved photoemission spectroscopy, we reveal that the top monolayer of ZrTe$_5$ crystals hosts a large band gap of ~100 meV on the surface and a finite constant density-of-states within the gap at the step edge. Our first-principles calculations confirm the topologically nontrivial nature of the edge states. These results demonstrate that the top monolayer of ZrTe$_5$ crystals is a large-gap 2D TI suitable for topotronic applications at high temperature.

# I. Introduction

Topological insulators (TIs) are a novel state of matter characterized by an energy gap in the bulk and gapless Dirac fermionic states connecting the valence and conduction bands on the boundary. For two-dimensional (2D) TIs, time-reversal symmetry (TRS) requires that electrons propagating in opposite directions have opposite spins in the one-dimensional (1D) edge state. Backscattering is prohibited as long as TRS is not broken by the scattering potential, leading to dissipationless transport edge channels and the quantum spin Hall (QSH) effect [1,2]. These exotic properties have been inducing fantastic reveries in electronic device applications. Although considerable theoretical efforts have predicted many 2D TI candidates in recent years [3-17], only a few systems, such as the quantum wells of HgTe/CdTe and InAs/GaSb, have been proved experimentally to be 2D TIs [18,19]. However, both of these hetero-junction materials display the QSH effect only at ultra-low temperatures due to a small bulk gap of the order of meV. Highly precise material control is required to fabricate such quantum well structures. These extreme requirements have obstructed their further studies and possible applications.

Recently, it has been proposed that $ZrTe_5$ monolayer is a 2D TI with a bulk gap of 100 meV [20]. Such a large gap is promising for the observation of the QSH effect and for device applications at room temperature. For a 2D TI, its three-dimensional (3D) stacked version could be either a 3D weak or strong TI, depending on the strength of the interlayer coupling [21]. First-principles calculations show that the $ZrTe_5$ crystal is located in the vicinity of a transition between strong and weak TI [20]. So far, over a dozen documented materials have been experimentally identified as strong TIs, which are characterized by an odd number of Dirac-cone-like in-gap states on each surface [Fig. 1(a)]. However, an unambiguous experimental confirmation of weak TIs is still elusive. For a weak TI stacked by 2D TI sheets, no in-gap surface state exists on the surface perpendicular to the stacking direction, and the side surface has an even number of Dirac-cone like surface states [Fig. 1(b)]. Further, in such a weak TI, when the top monolayer has a separated step edge, it is a 2D TI characterized by topologically nontrivial helical edge states as illustrated in Fig. 1(c).

In this work, by using scanning tunneling microscopy/spectroscopy (STM/S) and angle-resolved photoemission spectroscopy (ARPES), we systematically investigate the electronic states of the top surface and a step edge of ZrTe$_5$ crystal. We observe a large ~100 meV energy gap between the nearly linearly dispersive "cone-like" valence and conduction bands at the Brillouin zone center, in excellent agreement with the bulk band structure calculations. The STS spectrum shows zero density-of-states (DOS) inside the energy gap on the surface. More significantly, a finite constant DOS emerges inside the energy gap at the step edge, indicating the existence of linearly dispersive bands along the edge. The edge states are well reproduced by our single ribbon calculations, which show topologically nontrivial Dirac-like states inside the bulk gap at the step edge, thus providing strong evidences that the top monolayer of ZrTe$_5$ crystals is a large-gap 2D TI.

## II. Experimental and Computational Methods

Single crystals of ZrTe$_5$ were grown by the chemical vapor transport method. Stoichiometric amounts of Zr (powder, 99.2%, Hf nominal 4.5%) and Te (powder, 99.999%) were sealed in an evacuated quartz ampoule with iodine (7 mg/mL) and placed in a two-zone furnace. Typical temperature gradient from 480 °C to 400 °C was applied. After one month, long ribbon-shaped single crystals were obtained. STM measurements were carried out on ZrTe$_5$ single crystal samples that were cleaved *in situ* at ~20 K, then quickly inserted to the STM and cooled down to 4.3 K in ~3 mins. ARPES measurements were performed at the "Dreamline" beamline of the Shanghai Synchrotron Radiation Facility (SSRF) with a Scienta D80 analyzer. The energy and angular resolutions were set to 15–30 meV and 0.2°, respectively. The samples for ARPES measurements were cleaved *in situ* and measured in a temperature range between 24 and 200 K in a vacuum better than $5\times10^{-11}$ Torr.

The Vienna *ab initio* simulation package (VASP) [22,23] is employed for first-principles calculations. The generalized gradient approximation (GGA) of Perdew-Burke-Ernzerhof type [24] is used for the exchange-correlation potential. Spin-orbit coupling (SOC) is taken into account self-consistently. The cut-off energy for plane wave expansion is 500 eV and the *k*-point sampling grids for different structures have been tested to be dense enough. The atomic structure and the lattice constants *a*=3.987 Å, *b*= 14.502 Å, and *c*= 13.727 Å determined by power diffraction are adapted in our calculations [25]. The ZrTe$_5$ layers are stacked along the *b* axis. Band topology analysis indicates that increasing the interlayer distance by only 0.25 Å (~3.4%) will tune the material from strong TI to weak TI [20]. Such small change is within the error in determining the lattice constants by GGA calculations. In this work, we take slightly increased interlayer distance to mimic the weak TI phase, which is consistent with our experimental observations. A vacuum of 15 Å thick is used to minimize the interactions between the layers or ribbons and its periodic images. To largely reproduce the realistic (010) surface of the sample, we build a slab model of nine-unit-cell thick along the *b* axis [26,27]. To study the edge state along the *a* axis, a single ribbon of nine-unit-cell wide along the *c* axis, terminated by a $Te_2$ zigzag chain and a $ZrTe_3$ chain on each side, is calculated.

## III. Results

$ZrTe_5$ crystals have an orthorhombic layered structure characterized by space group *Cmcm*. As shown in Figs. 1(d) and 1(e), the triple prisms of $ZrTe_3$ form chains running along the *a* axis, and the adjacent chains are linked *via* parallel zigzag chains of Te atoms to form a 2D sheet of $ZrTe_5$ in the *a-c* plane with the lattice constants *a* = 3.987 Å and *c* = 13.727 Å. The sheets of $ZrTe_5$ stack along the *b* axis with a spacing of 7.251 Å, forming a layered structure. The adjacent $ZrTe_5$ layers are coupled *via* weak van der Waals forces, so that the cleavage mostly takes place between them.

Figure 1(f) displays a STM constant-current topographic image of a cleaved $ZrTe_5$ *a-c* (010) surface. We observe bright stripes parallel to each other with a separation of 13.9 Å, which matches the lattice constant *c*. The inset in the top right corner of Fig. 1(f) indicates that the stripes are constituted of dimers. The distance between the dimmers along the stripe direction is 4.1 Å, which matches the lattice constant *a*. The distance of the two bright spots in one dimer is 2.8 Å, which is consistent with that of the two Te atoms at the top of a triple prism. Thus, the bright stripes represent the $ZrTe_3$ chains in the termination layer. No obvious reconstruction is observed on the terminating surface. The electronic structure of the surface is spatially homogeneous on a clear surface. The inset in the bottom left corner of Fig. 1(f) shows a differential tunneling conductance spectrum, which is proportional to the DOS. The conductance drops to zero within an energy range of ~100 meV slightly above the Fermi energy ($E_F$), demonstrating the existence of a large energy gap in the DOS of the top layer.

To clarify the origin of the energy gap observed in the STS, we have performed systematic ARPES measurements on the (010) surface of $ZrTe_5$. Figure 2(a) shows that a hole-like band along Γ-X crosses $E_F$ in a "head-touching" style at low temperature, forming a small hole pocket at the BZ center, as shown in the Fermi surface intensity map in Fig. 2(e). The photon energy dependence measurements displayed in Fig. 2(d) reveal that the peaks of the momentum distribution curves exhibit a slight modulation along the $k_y$ direction with a period of $2\pi/b'$, where $b'$ is one half of the $b$-axis lattice constant of ZrTe$_5$, confirming the bulk nature of the measured electronic states. Due to the short escape depth of the photoelectrons excited by the vacuum-ultraviolet light used here, the present ARPES experiments mainly probe the electronic states of the top layer. This suggests that the top layer reflects the bulk electronic states of ZrTe$_5$ crystals, which are quasi-2D due to the weak interlayer coupling.

The experimental band dispersions are consistent with the calculated valence band structure. It is clear that $E_F$ lies slightly below the valence band top. This feature is consistent with the observation in the STM measurements, in which the conductance drops to zero approximately at a positive bias of 30 meV. To observe the conduction bands, we have attempted to deposit potassium atoms onto the surface, which can dope electrons into the surface and thus shift the chemical potential upwards. However, the chemical potential is only shifted to slightly above the valence band top, as shown in Fig. 2(c). Further deposition has no effect on the chemical potential. Another conventional method to observe the band structure slightly above $E_F$ is to broaden the Fermi-Dirac (FD) distribution function by raising temperature. Surprisingly, the chemical potential steadily shifts upwards with raising the sample temperature. This up-shifting of the chemical potential with increasing temperature is likely caused by the highly asymmetrical DOS across $E_F$, since a similar effect has been observed in Fe-based superconductors [28-30].

However, signals from the conduction bands are still unobserved in the pristine ZrTe$_5$ crystals in spite of the up-shifting of the chemical potential and the broadening of the FD distribution function at high temperature. We thus performed ARPES measurements at 190 K on the ZrTe$_5$ crystals with a small amount of Zr substituted by Ta, which dopes electrons slightly but changes the band dispersions little as compared to pristine ZrTe$_5$. The ARPES results of the substituted ZrTe$_5$ crystals recorded at 190 K are summarized in Fig. 3. From the intensity plot along Γ-X shown in Fig. 3(a), we can distinguish some weak signals above the hole-like band top. After division by the FD distribution function convoluted with the energy resolution function set in the ARPES measurements, the signals are much enhanced, as shown in Figs. 3(c) and 3(d). One peak is clearly observed at 60 meV above $E_F$ at the Γ point, corresponding to the conduction band bottom, which is consistent with the calculated band structure. Moreover, the ARPES data measured along Y-M exhibit very weak signals above $E_F$ at $k_x = 0.15$ Å$^{-1}$, which are attributed to the tail of the conduction band along Y-M, in agreement with the bulk band calculations [20]. This suggests that the band bottom along Y-M is a little higher than that at Γ. From the ARPES data, a global band gap of ~100 meV is determined and no Dirac-like band dispersion is observed inside the gap, which are consistent with the observations in the STS spectrum shown in the inset of Fig. 1(f).

According to the previous theoretical prediction [20], while the monolayer ZrTe$_5$ is a 2D TI, its 3D stacked compound should be either a weak or strong TI. The STS and ARPES spectra indicate that no surface state exists within the gap on the (010) surface and thus rules out the possibility of a strong TI. As discussed above, if the ZrTe$_5$ crystal is a weak TI stacked by 2D TI sheets, topologically nontrivial helical edge states should emerge inside the bulk gap at the step edge of a monolayer. Due to the weak inter-chain coupling in the *a-c* plane, the cleavage occurs not only between the adjacent ZrTe$_5$ layers but also between the ZrTe$_3$ and zigzag Te chains, leaving the step edges parallel to the chain direction along the *a* axis. To check whether the edge states exist, we have preformed STS measurements along a line crossing the step edge perpendicularly.

The inset of Fig. 4(a) shows the measured step topography and the locations of the data points registered with an interval of 1 nm. The step-height difference is measured to be 7.2 Å, which is consistent with the spacing between two adjacent ZrTe$_5$ layers, indicating that it is the edge of a single layer of ZrTe$_5$. In Fig. 4(a), the spectra away from the step edge on the lower (blue curves) and upper terraces (green curves) both show an energy gap of ~100 meV with zero conductance inside the gap. In contrast, the spectra around the step edge (red curves) have a finite conductance within the gap. To see the evolution of the in-gap states near the step edge more clearly, we zoom-in on the spectra taken from the 12$^{th}$ to 22$^{th}$ points in an energy range covering the energy gap and offset them with a constant incremental conductance in Fig. 4(c). The conductance becomes non-zero upon approaching the edge from both sides and reaches a maximum right at the edge (point #17), demonstrating the existence of in-gap edge states. Moreover, the non-zero in-gap conductance remains a constant over the energy gap range, indicating that the edge states have a constant DOS, which is usually associated with linear 1D band dispersions along the edge. Compared with STM measurements for bilayer Bi in the (111) plane [31-33], our results show direct and clear conducting edge states inside the energy gap of a top monolayer.

To illuminate the nature of the observed edge states, we have calculated the electronic states of a ribbon having left (terminated with $Te_2$ zigzag chain) and right (terminated with $ZrTe_3$ chain) step edges with and without the support of a thick $ZrTe_5$ slab. We find that the obtained edge states from the freestanding ribbon are nearly the same as the ones from the supported ribbon due to the weak interlayer coupling. The results for the freestanding ribbon are presented in Fig. 4(d). There is a Dirac-like edge state in the gap and the splitting in the lower branch is due to the nonequivalent left and right edges. The DOS from such edge states is shown in the inset of Fig. 4(b), which is nearly a constant finite value inside the gap region. Fig. 4(b) also shows the calculated local DOS from the top layer on a (010) slab with a thickness of nine layers, which can largely mimic the STM measurements far from the step edge, such as for point #1 and point #35. It exhibits a large energy gap of ~70 meV in the DOS and reproduces the experimental observations well. The excellent consistency of the results from the experiments and calculations indicates that the top monolayer of $ZrTe_5$ crystals is a large-gap 2D TI.

To specify the spatial distribution of the edge states, the normalized conductance integrals within the energy gap for individual spectra are plotted as a function of the distance from the edge and fitted with an exponential function, as shown in Fig 4(e). The exponential decay of the edge state, with a characteristic length $\xi$ = 1.5 nm, indicates that the edge state observed distributes closely along the step edge and extends laterally merely to a distance of the dimension of a primitive cell.

## IV. Discussion

The combined STM/S and ARPES results demonstrate that the top monolayer of $ZrTe_5$ crystals is a large-gap 2D TI, which can be used as a QSH device and ideal dissipationless wire operating at high temperature or even possibly at room temperature. This will largely stimulate the application of TIs and bring the birth of topotronics as the next generation of microelectronics and spintronics. Furthermore, the fabricated $ZrTe_5$ crystals with an insulating bulk gap inherently have many terraces and they can be viewed as a natural QSH device integrated with multiple channels. If an *s*-wave superconducting film covers such step edges, 1D topological superconductivity with Majorana bound states at the end of the terrace edges can be induced through the superconducting proximity effect. This will open up a new field for engineering topological quantum-computing devices.

## Acknowledgements

This work was supported by the Ministry of Science and Technology of China (No. 2015CB921300, No. 2013CB921700, and No. 2011CBA00108), the National Natural Science Foundation of China (No. 11227903, No. 11474340, No. 11422428, No. 11274362, and No. 11234014), the Chinese Academy of Sciences (No. XDB07000000), and the State of Texas through TcSUH. Partial calculations were preformed on TianHe-1(A), the National Supercomputer Center in Tianjin, China.

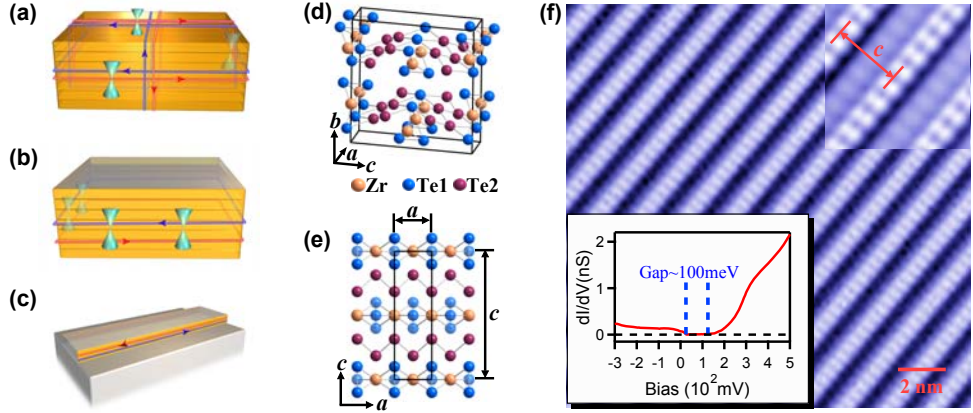

Fig. 1. (a) and (b) Schematics of 3D strong and weak TIs stacked by 2D TIs, respectively. Red and blue lines represent electrons propagating in opposite directions (indicated by arrows) with opposite spins in the topologically protected helical edge states on the boundary. (c) Schematic of topologically protected helical edge states existing along the step edge of a top monolayer of a weak TI stacked by 2D TI sheets. (d) 3D structure of $ZrTe_5$ crystal. (e) Top view of the single layer structure. (f) STM constant current topographic image of the cleaved $a$-$c$ (010) surface of $ZrTe_5$ crystal (I = 100 pA, V = -800 mV). Right top inset: High-resolution topography (I = 50 pA, V = 400mV) showing the details of the surface structure. Bottom left inset: tunneling differential conductance spectrum on the $a$-$c$ surface (I = 500 pA, V = 500 mV) showing a large energy gap of ~100meV above $E_F$ with zero-conductance inside the gap.

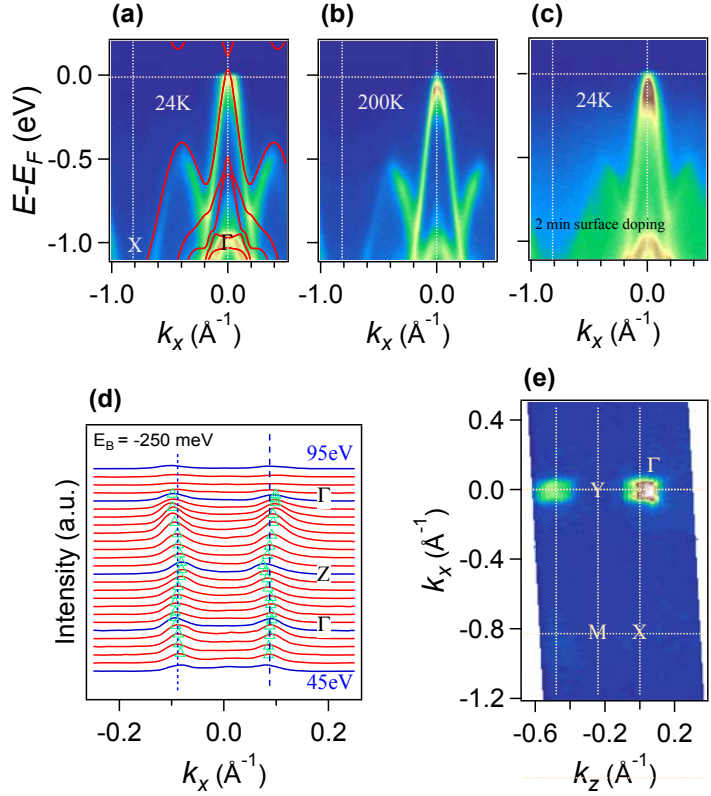

Fig. 2. Band structure of pristine ZrTe$_5$. (a) Band dispersions along Γ-X measured at 24 K. For comparison, the calculated band structure along Γ-X is plotted on top of the experimental data. The calculated bands are shifted up by 70 meV to match the experimental data. (b) Same as (a) but recorded at 200 K. (c) Same as (a) but measured after the deposition of potassium atoms for 2 minutes. (d) Momentum distribution curves at -250 meV recorded with different incident photon energies from 45 to 95 eV. (e) Fermi surface intensity plot recorded at 24 K showing a small hole pocket at Γ.

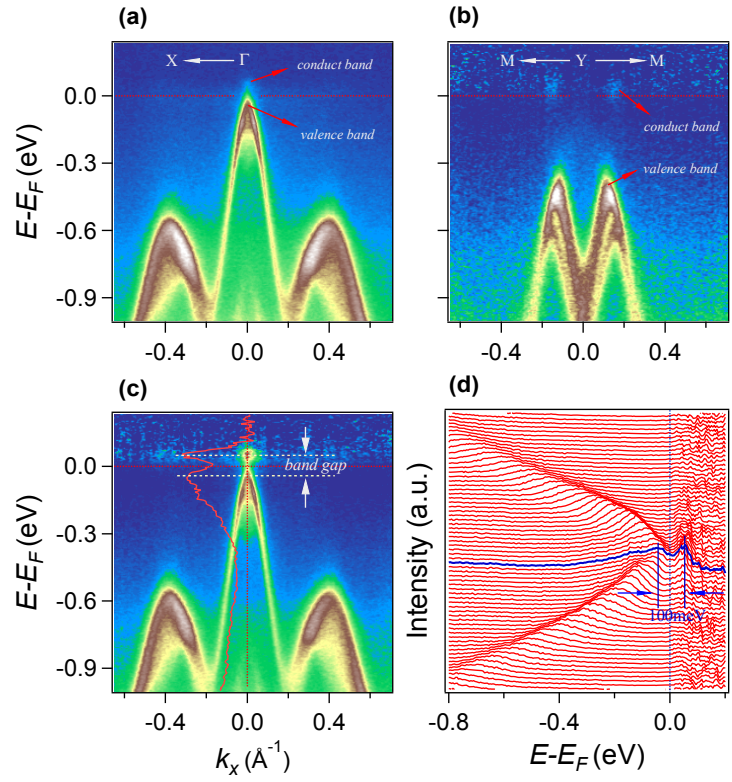

Fig. 3. Band structure of lightly-substituted $(Zr_{1-x}Ta_x)Te_5$ recorded at 190 K. (a) Intensity plot of the band dispersions along Γ-X. (b) and (c) Intensity plots of the band dispersions along Y-M and Γ-X, respectively, divided by the FD distribution function convoluted with the energy resolution function set in the ARPES measurements. The red curve in (c) is the energy distribution curve at Γ. (d) Energy distribution curve plot of (c).

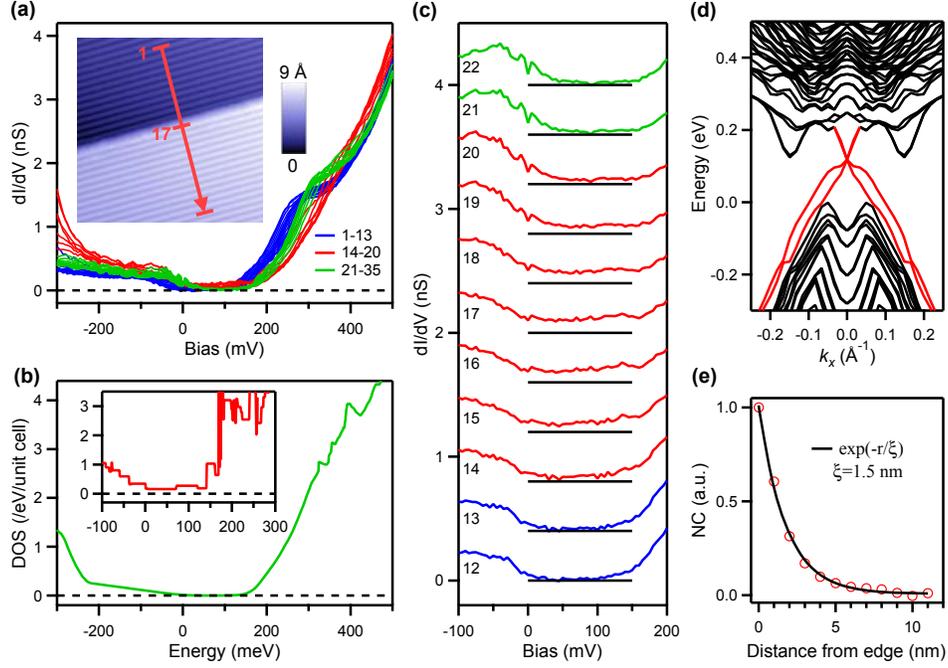

Fig. 4. (a) Tunneling differential conductance spectra along a line perpendicular to a monolayer step edge (I = 500 pA, V = 500 mV). Red curves are tunneling spectra measured near the step edge, while the blue/green ones are measured away from the step edge on the lower/upper terraces, respectively. The inset shows the topographic image of the step (I = 30 pA, V = 300 mV) and data point locations. (b) Calculated DOS of the top monolayer. The inset shows the calculated DOS of the edge states of the monolayer along the chain direction. (c) Evolution of the edge states. For clarify, the spectra from the 12$^{th}$ to 22$^{th}$ points are offset with equal interval of 0.4 nS. Black lines indicate the zero-conductance level for each spectrum. (d) Calculated band dispersions of the edge states of the monolayer along the chain direction. (e) Normalized conductance (NC) integral within the gap plotted as a function of the distance from the edge. The black curve is an exponential function fit to the data with a characteristic decay length $\xi$ = 1.5 nm.